\documentclass[inorganics,article,submit,moreauthors,pdftex,10pt,a4paper]{mdpi}

\firstpage{1} 
\articlenumber{x}
\pdfoutput=1

\Title{Single crystal growth and anisotropic magnetic properties of
Li$_2$Sr[Li$_{1-x}$Fe$_x$N]$_2$}

\Author{Peter H\"ohn$^{ 1}$, Tanita J. Ball$\acute{\rm e}^{ 2}$, Manuel Fix$^{ 2}$, Yurii Prots$^1$ and Anton Jesche$^{ 2,}$*}
\AuthorNames{Peter H\"ohn, Tanita J. Balle and Anton Jesche}

\address{%
$^{1}$ \quad Max-Planck-Institut f\"ur Chemische Physik fester Stoffe, N\"othnitzer Str. 40, D-01187 Dresden, Germany\\
$^{2}$ \quad EP VI, Center for Electronic Correlations and Magnetism, Augsburg University, D-86159 Augsburg, Germany}

\corres{Correspondence: anton.jesche@physik.uni-augsburg.de; Tel.: +49-821-598-3659}

\abstract{{Up to now, investigation of physical properties of ternary and higher nitridometalates was severely hampered by challenges concerning phase purity and crystal size. 
Employing a modified lithium flux technique, we are now able to prepare sufficiently large single crystals of the highly air and moisture sensitive nitridoferrate Li$_2$Sr[Li$_{1-x}$Fe$_x$N]$_2$ for anisotropic magnetization measurements. 
The magnetic properties are most remarkable: large anisotropy and coercivity fields of 7 Tesla at $T = 2$\,K indicate a significant orbital contribution to the magnetic moment of iron. 
Altogether, the novel growth method opens a route towards interesting phases in the comparatively recent research field of nitridometalates and should be applicable to various other materials.
}}


\begin{document}
\nolinenumbers
\section{Introduction}
Binary transition metal nitrides attract considerable interest due to their valuable mechanical, electrical and magnetic properties. In contrast, chemistry and physics of multinary nitrides are far less explored\,\cite{Kniep2013}. Nitridometalates of $d$ metals $T$ represent an interesting class of solid state phases, which contain nitrogen as isolated anions N$^{3-}$ or feature complex anions [$T_x^{}$N$_y$]$^{z-}$ of different dimensionalities with coordination numbers of $T$ by N typically between two and four and oxidation states of the transition metals comparatively low. Whereas the bonding within these complex anions and frameworks is essentially covalent, nitridometalates are stabilized by predominantly ionic bonding through counterions of electropositive metals like alkali ($A$) or alkaline-earth ($AE$) cations. 

%The introduction should briefly place the study in a broad context and highlight why it is important. It should define the purpose of the work and its significance. The current state of the research field should be reviewed carefully and key publications cited. Please highlight controversial and diverging hypotheses when necessary. 
Iron in nitridoferrates of alkali and alkaline-earth metals may be coordinated tetrahedrally (Li$_3$[Fe$^{\rm III}$N$_{4/2}$]\,\cite{Gudat1990, Nishijima1994})
 or trigonal-planar in isolated units ((Ca$_3$N)$_2$[Fe$^{\rm III}$N$_3$]\,\cite{Cordier1990}, Sr$_3$[Fe$^{\rm III}$N$_3$]\,\cite{Bendyna2008}, Ba$_3$[Fe$^{\rm III}$N$_3$]\,\cite{Hohn1991b}, (Sr$_{1-x}$Ba$_x$)$_3$[Fe$^{\rm III}$N$_3$]\,\cite{Hohn1992b}, Sr$_8$[Fe$^{\rm III}$N$_3$][Fe$^{\rm II}$N$_2$]\,\cite{Bendyna2008b, Bendyna2009}) 
as well as in oligomers (Ca$_2$[Fe$^{\rm II}$N$_2$]\,\cite{Hohn1992c}, Sr$_2$[Fe$^{\rm II}$N$_2$]\,\cite{Hohn1992c}) and 1D chains (LiSr$_2$[Fe$^{\rm II}_2$N$_3$]\,\cite{Hohn1991c}, LiBa$_2$[Fe$^{\rm II}_2$N$_3$]\,\cite{Hohn1991c}).
Linear coordination is observed as linear dumbbells in Sr$_8$[Fe$^{\rm III}$N$_3$][Fe$^{\rm II}$N$_2$]\,\cite{Bendyna2008b, Bendyna2009}, Sr$_8$[Mn$^{\rm III}$N$_3$][Fe$^{\rm II}$N$_2$]\,\cite{Bendyna2008c}, Sr$_2$[Fe$^{\rm II}$N$_2$]\,\cite{Hohn1992c}, and Li$_4$[Fe$^{\rm II}$N$_2$]\,\cite{Gudat1991}, 
and in linear substituted chains [(Li$_{1-x}$Fe$^{\rm I}_x$)N]$^{2-}$ in 
Li$_2$[(Li$_{1-x}$Fe$^{\rm I}_x$)N]\,\cite{Niewa2003, Niewa2003b, Klatyk2002}, Li$_2$Ca[(Li$_{1-x}$Fe$^{\rm I}_x$)N]$_2$\,\cite{Klatyk1999b}, Li$_2$Sr[(Li$_{1-x}$Fe$^{\rm I}_x$)N]$_2$\,\cite{Klatyk1999b},
LiCa$_2$[(Fe$^{\rm I}_{1-x}$Li$_x$)N$_2$]\,\cite{Klatyk1999c} and
LiSr$_2$[(Fe$^{\rm I}_{1-x}$Li$_x$)N$_2$]\,\cite{Hohn2014}. 
Structural data for the majority of nitridoferrates reported up to now were derived from single crystal data, whereas in most cases single phase powder samples had to be employed for the investigation of physical properties, with the exception of Li$_2$[(Li$_{1-x}$Fe$^{\rm I}_x$)N]\,\cite{Jesche2014c} and LiSr$_2$[(Fe$^{\rm I}_{1-x}$Li$_x$)N$_2$]\,\cite{Hohn2014}, where single crystals of sufficient size were available for physical properties investigations.

The single crystal growth of nitrides is often challenging due to the large dissociation energy of the N$_2$ molecule, the reactivity of the starting materials and enhanced vapor pressures. 
Only recently, Li-rich flux was successfully used for the growth of large single crystals of LiCaN, Li$_3$N and Li$_2$(Li$_{1-x}T_x$)N with $T = \rm{Mn, Fe, Co}$\,\cite{Jesche2014c} and $T = {\rm Ni}$\,\cite{Jesche2015}. 
%Whether the use of this method can be extended to...
%Whether this method is applicable to more complex systems, in particular ternary or quaternary nitrides that contain alkaline earth elements, has not been settled.  
Further development of the high temperature centrifugation aided filtration technique\,\cite{Fisk1989, Canfield1992, Bostrom2001} by addition of Na and NaN$_3$ to increase the basicity of the flux also enabled the growth of large single crystals of nitride metalides like Li$_{16}$Sr$_6$Ge$_6$N\,\cite{fehlt}. 
However, the extent of application of this method towards more complex systems, in particular transition metal rich ternary or quaternary nitridometalates that also contain alkaline earth elements, has not been investigated in detail.
Magnetic properties were reported only for few nitridometalates containing alkaline-earth metals. 
Some of these phases show ferromagnetic (LiSr$_2$[CoN$_2$]\,\cite{Hohn2014}) or antiferromagnetic (LiSr$_2$[FeN$_2$]\,\cite{Hohn2014}) ordering. Furthermore, for many phases (for example Sr$_8$[MnN$_3$]$_2$[MnN$_2$]\,\cite{Ovchinnikov2015}) a large spin-orbit coupling together with low or intermediate spin states is being discussed. 

\begin{figure}
\centering
\includegraphics[width=0.9\textwidth]{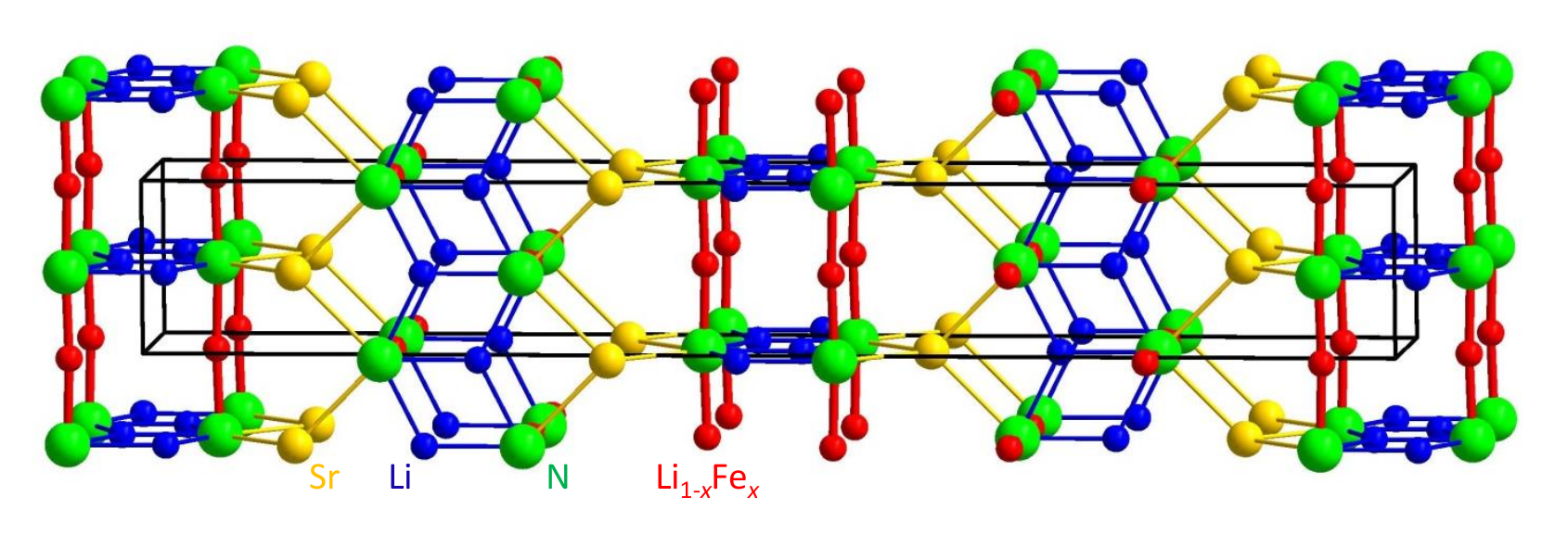}
\caption{Crystal structure of Li$_2$Sr[(Li$_{1-x}$Fe$_x$)N]$_2$, the unit cell is indicated by the black lines (space group $I4_1\!/\!amd$). Fe is substituted in linear, two-fold coordination between N.}
\label{structure}
\end{figure}

We shall present results on Fe-substituted Li$_4$SrN$_2$. 
The peculiar (almost) linear, two-fold coordination of Fe makes this material particularly promising since unprecedented magnetic coercivity and anisotropy were found in Fe-substituted Li$_3$N, which shares the same structural feature\,\cite{Klatyk2002, Jesche2014b}. 
Synthesis and crystal structure of Fe-substituted Li$_4$SrN$_2$ was first reported by Klatyk and Kniep\,\cite{Klatyk1999b}: 
Small single crystals sufficient for X-Ray diffraction were obtained by reaction of Li, Li$_2$(Li$_{0.66}$Fe$_{0.33}$)N and Sr$_2$N in a molar ratio of 7:6:4. 
The compound crystallizes in a tetragonal lattice, space group $I4_1\!/\!amd$ (No. 141) with $a = 3.7909(2)$\,\AA, and $c = 27.719(3)$\,\AA. 
As indicated by the notation of the chemical formula, Li$_2$Sr[(Li$_{1-x}$Fe$_x$)N]$_2$ with $x = 0.46$, the substituted Fe atoms occupy only one of the two Li-sites (the one in two-fold coordination of N, see Fig.\,\ref{structure}).
The Fe-coordination is not strictly linear. 
Rather the N-Fe-N angle amounts to 177.4$^\circ$ as inferred from the reported crystal structure\,\cite{Klatyk1999b}. 
The Fe-substitution causes a decrease of $a$ but an increase of $c$ by 0.8\,\% and 2.5\,\%, respectively, compared to the parent compound Li$_4$SrN$_2$\,\cite{Cordier1989b}. 
No physical properties have been reported so far.
%whereas the second site is 

%This is a figure, Schemes follow the same formatting. If there are multiple panels, they should be listed as: (\textbf{a}) Description of what is contained in the first panel. (\textbf{b}) Description of what is contained in the second panel. Figures should be placed in the main text near to the first time they are cited. A caption on a single line should be centered.

%Finally, briefly mention the main aim of the work and highlight the principal conclusions. As far as possible, please keep the introduction comprehensible to scientists outside your particular field of research. Citing a journal paper 
Here we show that Li$_2$Sr[(Li$_{1-x}$Fe$_x$)N]$_2$ single crystals of several millimeter along a side can be obtained from Li-rich flux. 
The large magnetic anisotropy and coercivity revealed a significant orbital contribution to the magnetic moment of Fe. 
Cu-substituted Li$_4$SrN$_2$ was investigated as a non-local-moment-bearing reference compound.

%%%%%%%%%%%%%%%%%%%%%%%%%%%%%%%%%%%%%%%%%%
\section{Results}

%This section may be divided by subheadings. It should provide a concise and precise description of the experimental results, their interpretation as well as the experimental conclusions that can be drawn.

%%%%%%%%%%%%%%%%%%%%%%%%%%%%%%%%%%%%%%%%%%
\subsection{Single crystal growth}
\begin{figure}
\centering
\includegraphics[width=0.9\textwidth]{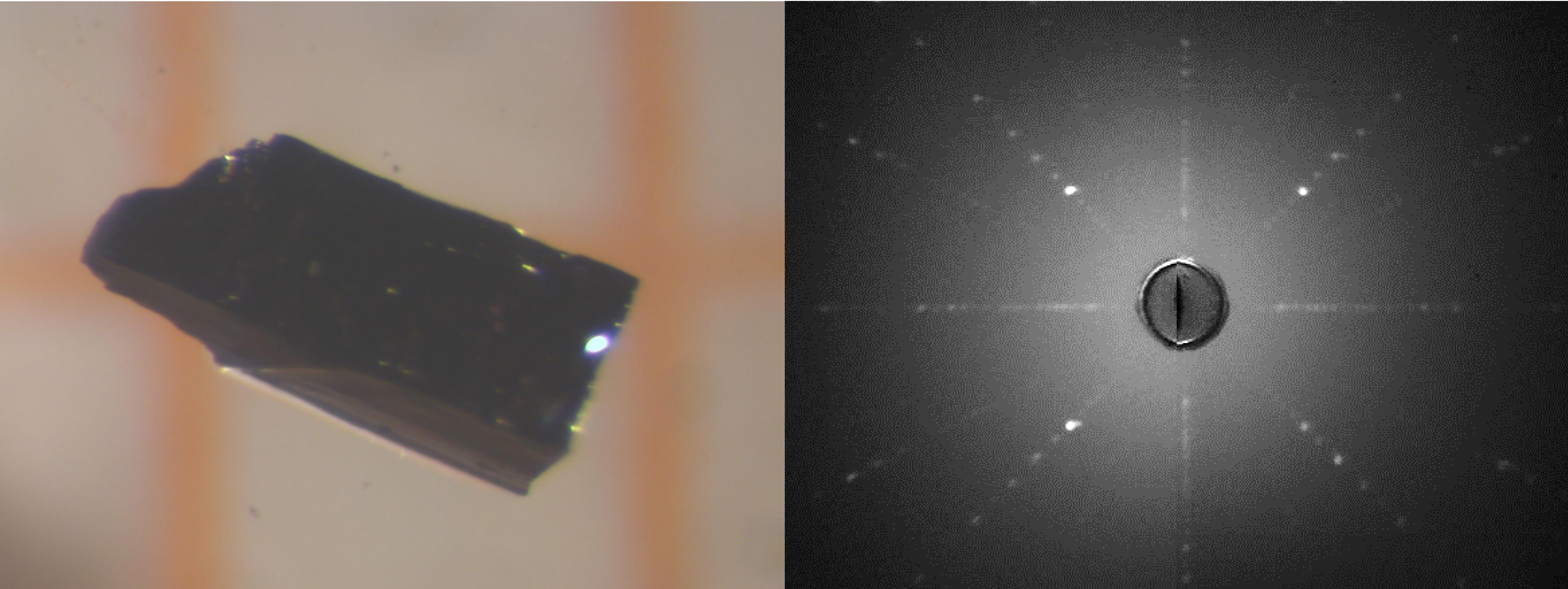}
\caption{Li$_2$Sr[(Li$_{1-x}$Fe$_x$)N]$_2$ single crystal ($x = 0.41$) on a millimeter grid and corresponding Laue-back-reflection pattern to the right showing the four-fold rotational symmetry along the crystallographic $c$-axis. }
\label{kristall}
\end{figure}

Due to the air and moisture sensitivity of both the reactants (Li, Sr$_2$N, NaN$_3$) and the final product Li$_2$Sr[(Li$_{1-x}$Fe$_x$)N]$_2$ all manipulations including grinding and weighing, as well as  complete sample preparations for measurements were carried out in an inert gas glove box (Ar, O$_2$ and H$_2$O $\leq 1$\,ppm).
The title compound was obtained in form of large, black single crystals from the reaction of Sr$_2$N, Fe, Li and NaN$_3$ in molar ratio $1 : 1.8 : 36 : 1$ with NaN$_3$ acting as nitrogen source and Li acting as flux and mineralizer.  
The mixtures with a total mass of roughly 0.6\,g were placed in a tantalum ampule\,\cite{Canfield2001}.
The whole device was sealed by arc welding under inert atmosphere of 700\,mbar argon and subsequently encapsulated in a quartz tube with an internal argon pressure of 300\,mbar in order to prevent oxidization of the tantalum. 
The sample was heated from room temperature to $T = 700\,^\circ$C within 7\,h, annealed for 2\,h, then cooled to $T = 300\,^\circ$C within 400\,h, and finally centrifuged with 3000\,min$^{-1}$ to separate the single crystals of a blackish color from the excess flux. 
Besides several large single crystals of the title phase, small amounts of single crystalline Li$_3$N and LiSr$_2$[(Fe$_{1-x}$Li$_x$)N$_2$]\,\cite{Hohn2014} were also obtained. 
Using a two-probe multimeter, the different phases showed remarkably different resisitivities enabling the discrimination of the various phases. 
%The starting materials were mixed in a molar ratio of Li:Li$_3$N:Fe:Sr = X:X:X:X. The mixtures with a total mass of roughly XX\,g were packed in a three-cap Ta crucible.

A representative Li$_2$Sr[(Li$_{1-x}$Fe$_x$)N]$_2$ single crystal is shown in Fig.\,\ref{kristall}. 
The samples show a plate-like habit with the crystallographic $c$-axis oriented perpendicular to the large surface as confirmed by Laue-back-reflection (right panel in Fig.\,\ref{kristall}). The spot-size of the X-Ray beam was similar to the sample size.

An Fe-concentration of $x = 0.42$ was determined by energy-dispersive X-Ray analysis (EDX) based on the Fe:Sr ratio and assuming fully occupied Sr-sites. 
An almost identical value of $x = 0.41$ was found by chemical analysis by means of inductively coupled plasma optical emission spectroscopy (ICP-OES; accessible is the Li:Sr:Fe ratio) and was confirmed with a second sample taken from the same batch.
The observed slight Li-excess of 0.15 per formula unit (0.19 for the other sample), is attributed to small amounts of Li-rich flux remnants.

\subsection{Crystal structure}

\begin{figure}
\centering
\includegraphics[width=0.8\textwidth]{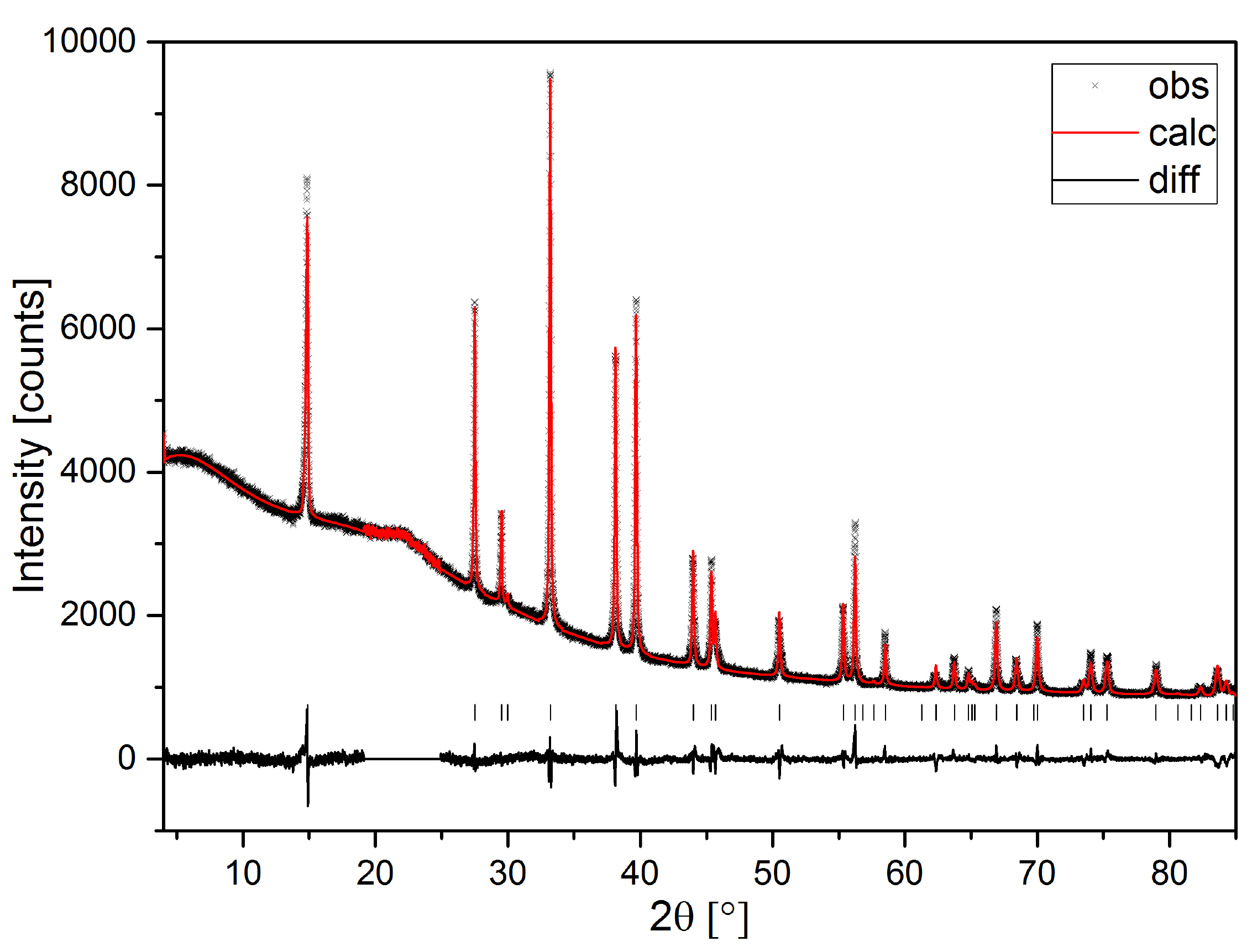}
\caption{X-Ray diffraction pattern measured on ground single crystals of Li$_2$Sr[(Li$_{1-x}$Fe$_x$)N]$_2$ (Co $K \alpha $ radiation, $\lambda =  1.78892$\,\AA). Tick marks correspond to reflection positions of Li$_2$Sr[(Li$_{1-x}$Fe$_x$)N]$_2$.}
\label{xrpd}
\end{figure}

\begin{table}
\caption{Single crystal structure refinement for Li$_2$Sr[(Li$_{1-x}$Fe$_x$)N]$_2$.}
\centering
\begin{tabular}{ll}
\toprule
\midrule
crystal system													& tetragonal				\\
space group														& $I4_1\!/\!amd$ (No. 141)	\\
$a$ (\AA)														& 3.8011(1)				\\
$c$	(\AA)														& 27.586(3)					\\ 
Fe occupancy													& 0.32(1)						\\
cell volume (\AA$^3$)											& 398.57(5)					\\
$Z$																& 4							\\
$\rho_{\rm calcd}$ (gcm$^{-3}$)										& 2.907 					\\
crystal color, habit											& black, tetragonal column	\\
crystal size (mm)													& 0.02 x 0.02 x 0.07		\\
$\mu$(Mo$_{K\alpha}$, mm$^{-1}$)											& 15.52						\\
2\textit{$\theta$} range ($^{\circ}$)							& 5.90 - 59.60 				\\
diffractometer													& RIGAKU					\\
wavelength (\AA)					 							& 0.71069 (Mo $K\,\alpha$)  \\
monochromator													& graphite					\\
temperature														& 293 K						\\
scan mode														& profile data from $\phi$ scans						\\
measured reflections											& 3355						\\
independent reflections											& 187						\\
observed reflections [\textit{F}$_{\rm o}$ > 4$\sigma$(\textit{F}$_{\rm o}$)]	& 2919  \\
\textit{R}$_{\rm int}$												& 0.041						\\
structure solution method										& direct					\\
number of parameters											& 17						\\
goodness-of-fit on \textit{F}$^{2}$								& 1.081						\\
\textit{wR}2																	& 0.078     \\
\textit{R}1 [\textit{F}$_{\rm o}$ > 4$\sigma$(\textit{F}$_{\rm o}$)]		& 0.030	    \\
\textit{R}1 (all data)															& 0.035  	\\
residual electron density (e$\times 10^{-6}$ pm$^{-3}$)							& 1.24/-1.70\\
\bottomrule
\end{tabular}
\label{tab}
\end{table} 

Figure\,\ref{xrpd} shows the X-Ray powder diffraction pattern measured on ground Li$_2$Sr[(Li$_{1-x}$Fe$_x$)N]$_2$ single crystals.
No foreign phases were detected. The region between $2\theta = 20^\circ$-\,$25^\circ$ was excluded due to amorphous constituents that are created by degradation during the measurement.
Lattice parameters of $a = 3.79536(9)$\,\AA~and $c = 27.6492(13) $\,\AA~and an Fe occupancy of $x = 0.32$ were obtained by Rietveld refinement. 
The decrease in $a$ and increase in $c$ in comparison to the parent compound Li$_4$SrN$_2$ ($a = 3.822(2)$\,\AA~and $c = 27.042(9)$\,\AA\,\cite{Cordier1989b}) are slightly smaller than reported in Ref.\,\cite{Klatyk1999b} ($x = 0.46, a = 3.7909(2)$\,\AA~and $c = 27.719(3)$\,\AA) in accordance with a somewhat lower Fe concentration. 

Smaller crystals were obtained from crushed samples and selected for single crystal X-Ray diffraction.
The results are summarized in Table\,\ref{tab}.
Powder as well as single crystal X-Ray diffraction revealed an Fe-concentration of $x = 0.32$. This is somewhat smaller than the values obtained by EDX and chemical analysis and indicates that the Fe concentration may vary between different samples; the slight difference between powder and single crystal data may also stem from slight inhomogeneities within the samples. 
The single crystal used for the magnetization measurements (see below) was directly analyzed by ICP-OES and showed an iron concentration of $x = 0.41$. 

\subsection{Magnetic properties}
\begin{figure}
\centering
\includegraphics[width=0.9\textwidth]{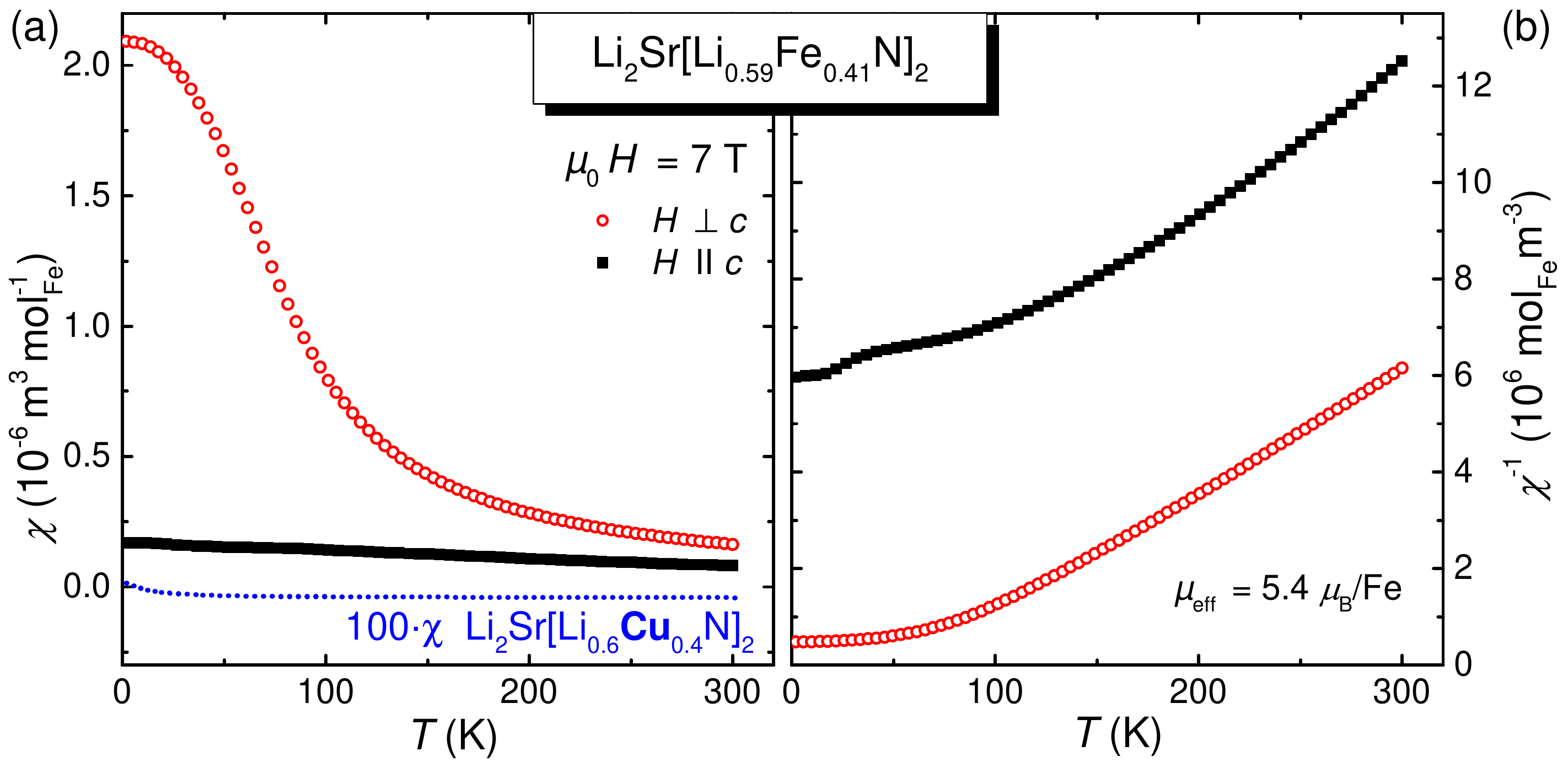}
\caption{Magnetic susceptibility $\chi = M/H$ per mol Fe as a function of temperature for field applied parallel and perpendicular to the crystallographic $c$-axis. (\textbf{a}) A pronounced anisotropy with larger $\chi$ for $H \perp c$ (open, red symbols) is observed up to room temperature. The molar susceptibility of Li$_2$Sr[(Li$_{0.6}$Cu$_{0.4}$)N]$_2$ multiplied by a factor of 100 is shown for comparison (blue, dotted line, $H \perp c$).
(\textbf{b}) The temperature dependence of the inverse susceptibility roughly follows a Curie-Weiss law for $T > 150$\,K.}
\label{chi}
\end{figure} 

The temperature-dependent magnetic susceptibility $\chi(T) = M/H$ measured parallel and perpendicular to the crystallographic c-axis is shown in Figure
\,\ref{chi}a (field-cooled in $\mu_0H = 7$\,T).
A temperature-independent contribution of a ferromagnetic impurity phase (with a Curie temperature significantly above room temperature, presumably elemental Fe or Fe$_3$O$_4$) was subtracted by assuming that the intrinsic, local-moment contribution of the title compound is linear in field at $T = 300$\,K (analogous to the Honda-Owen method).
%The value was estimated from isothermal magnetization measurements at room temperature. For both orientations the signal amounts to 
A pronounced anisotropy is observed over the whole temperature range investigated.
The ratio of $\chi_{\perp c}/\chi_{\parallel c}$ increases upon cooling from a value of 2 at $T = 300$\,K to 12 at $T = 2$\,K.

In order to confirm the proposed, unusual valence state of Fe(I)\,\cite{Klatyk1999b}, we performed further magnetization measurements on Li$_2$Sr[(Li$_{0.6}$Cu$_{0.4}$)N]$_2$ (the sample was grown similar to the isotypic title compound Li$_2$Sr[(Li$_{1-x}$Fe$_x$)N]$_2$).
The local moment behavior associated with the spin-1/2 of Cu(II) is supposed to be markedly different from the one of Cu(I).
As shown by the blue, dotted line in Fig.\,\ref{chi}a, the susceptibility of Li$_2$Sr[(Li$_{0.6}$Cu$_{0.4}$)N]$_2$ does not show local moment behavior and is indeed negligibly small compared to the one of the Fe-substituted homologue.
The largely temperature independent value of $\chi = -5(1)\cdot 10^{-10}\rm{m}^3\,{\rm mol}^{-1}$ is in reasonable agreement with the ionic diamagnetic contribution of the Li$_4$SrN$_2$ host material of $\chi = -6\cdot 10^{-10}\rm{m}^3\,{\rm mol}^{-1}$ 
\{assuming $\chi(\rm Li^{1+}) = -0.2\cdot10^{-10}\rm{m}^3\,{\rm mol}^{-1}$\,\cite{Banhart1986},  
$\chi(\rm Sr^{2+}) = -2\cdot10^{-10}\rm{m}^3\,{\rm mol}^{-1}$\,\cite{Myers1952} and  
$\chi(\rm N^{3-}) = -1.6\cdot10^{-10}\rm{m}^3\,{\rm mol}^{-1}$\,\cite{Hohn2009}\}.
Accordingly, the observed non-local moment behavior of Li$_2$Sr[(Li$_{1-x}$Cu$_x$)N]$_2$ implies the presence of Cu(I) and supports a valence state of Fe(I).
 
The inverse susceptibility roughly follows a Curie-Weiss law for $T = 150$\,K\,-\,300\,K (Figure\,\ref{chi}b).
For $H \perp c$ an effective moment of $\mu_{\rm eff} = 5.4\,\mu_{\rm B}$ per Fe and a ferromagnetic Weiss temperature of $\Theta_{\rm W} = 49$\,K were obtained.
The fit to the data considered a minor diamagnetic contribution of $\chi_0 = -1.5 \cdot 10^{-8}\rm{m}^3\,{\rm mol}^{-1}$ (10\% of the absolute value at room temperature).
The slope of $\chi^{-1}$ suggests a similar value of the effective moment for $H \parallel c$, however, the small absolute value of $\chi$ in combination with a large antiferromagnetic Weiss temperature prohibits an accurate estimate. 

\begin{figure}
\centering
\includegraphics[width=0.9\textwidth]{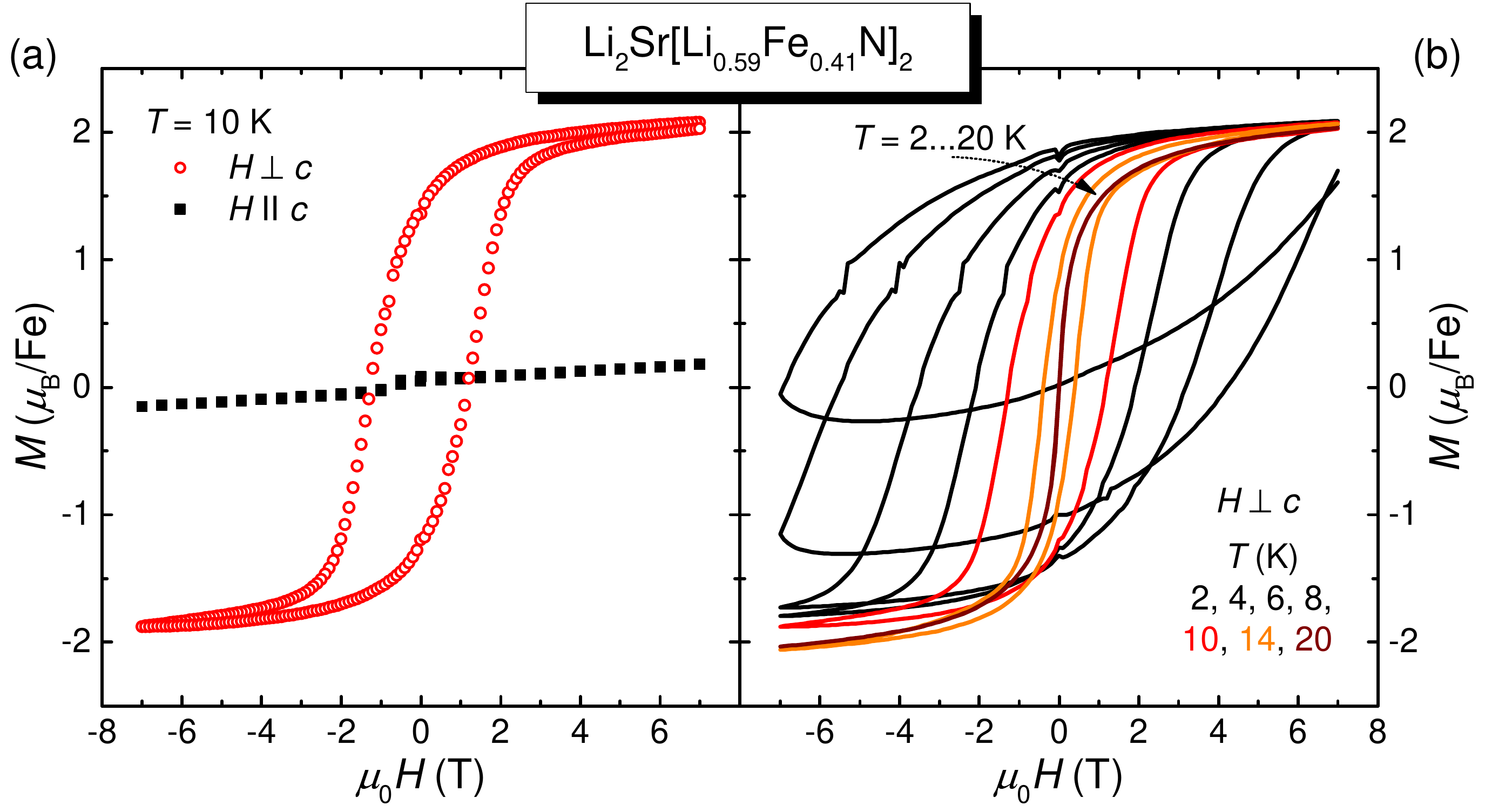}
\caption{Isothermal magnetization in $\mu_{\rm B}$ per Fe at $T = 10$\,K. (\textbf{a}) Large hysteresis emerges for field applied perpendicular to the $c$-axis (which is perpendicular to the N-Fe-N 'molecular axis'). (\textbf{b}) The hysteresis for $H \perp c$ increases rapidly with decreasing temperature. Spontaneous magnetization disappears for temperatures larger than $T \sim 16$\,K. 
\label{m-h}}
\end{figure}

The isothermal magnetization in $\mu_{\rm B}$ per Fe as a function of an applied magnetic field at $T = 10$\,K is shown in Figure\,\ref{m-h}a. 
For $H \parallel c$ the magnetization increases slowly with the applied field in a linear fashion without any appreciable hysteresis.
The magnetization is significantly larger for $H \perp c$ and exceeds values of 2\,$\mu_{\rm B}$ per Fe.
However, no saturation is observed even at the largest available field of $\mu_0H = 7$\,T.
In accordance with the large anisotropy, a pronounced hysteresis loop with a coercive field of $\mu_0 H_c = 1.2$\,T forms for $H \perp c$. 
The coercivity field increases rapidly upon cooling (Fig.\,\ref{m-h}b). 
At the lowest accessible temperature $T = 2$\,K the coercivity reaches almost 7\,T. 
The hysteresis vanishes for temperatures higher than $T \approx 16$\,K.
Furthermore, the $M-H$ loops are asymmetric for $T < 14$\,K: The (field cooled) value in $\mu_0H = +7$\,T is larger than the corresponding value found at $\mu_0H = -7$\,T.

The small anomalies at $H \sim 0$ and at $M \sim 1\,\mu_{\rm B}$ result from the subtraction of the ferromagnetic impurity contribution (which is not fully temperature independent) and a zero-crossing of the raw data signal, respectively. 
A significant in-plane anisotropy for different orientations perpendicular to the $c$-axis could be present, in particular [1\,0\,0] vs. [1\,1\,0] (see discussion). 
The closer investigation of this anisotropy, however, is beyond the scope of this publication.

%%%%%%%%%%%%%%%%%%%%%%%%%%%%%%%%%%%%%%%%%%
\section{Discussion}
The successful use of a Li-rich flux for the growth of large single crystals of Fe-substituted Li$_4$SrN$_2$ has been anticipated since there are only a few binaries known that may compete with the formation of this compound. Among those, Li-Sr binaries are not considered exceedingly stable as indicated by their low peritectic decomposition temperatures ($< 200^\circ$C)\,\cite{Massalski1996}.
Nevertheless, a good solubility of Sr in Li is inferred.
Only a few Sr-N binary compounds are known and non of those seem to be stable in the presence of Li.
No binary compounds of Sr-Fe or Sr-Ta (which may prevent the use of Ta crucibles) are known.
Furthermore, our work shows that Li$_3$N is not so stable that it prevents the formation of other nitrides.
Whether the related compounds LiSr$_2$Fe$_2$N$_3$\,\cite{Hohn1991c} or LiSr$_2$FeN$_2$\,\cite{Hohn2014} (isotypic to LiSr$_2$CoN$_2$\,\cite{Hohn1992}) can be grown as large single crystals by adjusting temperature profile and/or ratio of the starting materials is subject of ongoing research. 
The situation is similar for ternary and multinary nitrides that contain other alkaline earth and/or other transition metals and it seems not unlikely to find a wide range of applications for the Li-flux method.

The magnetic properties of Li$_2$Sr[(Li$_{1-x}$Fe$_x$)N]$_2$, in particular the large anisotropy and coercivity, are most remarkably. 
There are only a few materials known that show coercivity fields in the range of $\mu_0H_c \sim 7$\,T or above. 
Those are melt-spun ribbons of rare-earth-based Dy-Fe-B and Tb-Fe-B alloys (reported are $\mu_0H_c = 6.4$\,T for both materials\,\cite{Pinkerton1986} and $\mu_0H_c = 7.7$\,T for the latter one\,\cite{Liu2011}). 
For transition-metal-based compounds, the only examples we are aware of are LuFe$_2$O$_4$ ($\mu_0H_c = 9$\,T\,\cite{Wu2008} and $\mu_0H_c = 11$\,T\,\cite{Iida1993}) and Fe-substituted Li$_3$N ($\mu_0H_c = 11.6$\,T\,\cite{Jesche2014b}).
The large magnetic anisotropy and coercivity of both materials is caused by a significant orbital contribution to the magnetic moment of Fe (see \cite{Ko2009} for the former and \cite{Novak2002, Antropov2014, Jesche2015, Ke2015} for the latter one). 
Whereas the emergence of the orbital moment in LuFe$_2$O$_4$ is a result of a complex interplay between charge ordering, ferroelectricity and ferrimagnetism\,\cite{Yang2015}, the orbital moment in Fe-substituted Li$_3$N seems to be directly linked to the linear, two-fold coordination of Fe\,\cite{Jesche2014b}. 
Such a linear molecule or linear chain is not subject to a Jahn-Teller distortion\,\cite{Jahn1937, Kugel1982} which is the driving force for the quenching of the orbital magnetic moment that is usually observed in transition metals. 

As found for Fe-substituted Li$_3$N, the magnetic hard axis of Li$_2$Sr[(Li$_{1-x}$Fe$_x$)N]$_2$ is oriented perpendicular to the N-Fe-N 'molecular axis' (see Figures\,\ref{structure} and \,\ref{m-h}).
Accordingly, there is no unique easy-axis present in Li$_2$Sr[(Li$_{1-x}$Fe$_x$)N]$_2$ since the N-Fe-N molecules run along both the $a$- and the $b$-axes (in contrast to Fe-substituted Li$_3$N where the N-Fe-N molecules do define the easy axis and are oriented along the unique, hexagonal $c$-axis). 
The magnetization in the $a$-$b$ plane is therefore expected to reach between 1/2 and 1/$\sqrt 2$ of the saturation magnetization of Fe corresponding to field along $\langle 1\,0\,0\rangle$ and $\langle 1\,1\,0\rangle$, respectively. 
The increase of the magnetization for $H \perp c$ is only slightly larger than for $H \parallel c$ (see Fig.\,\ref{m-h}a).
This implies similar large magnetic anisotropy energies for $\langle 1\,1\,0\rangle$ and $\langle 0\,0\,1\rangle$ (with respect to $\langle 1\,0\,0\rangle$).

To summarize, the large magnetic anisotropy and coercivity observed in Li$_2$Sr[(Li$_{1-x}$Fe$_x$)N]$_2$ give strong evidence for a significant orbital contribution to the magnetic moment of Fe. The similarities to Li$_2$(Li$_{1-x}$Fe$_x$)N\,\cite{Jesche2014b, Jesche2015} are apparent. 
The origin of this behavior is attributed to the (almost) linear, two-fold coordination Fe. 
Our work shows that small deviations from linearity, that is a N-Fe-N angle of $\sim$177$^\circ$ instead of $180^\circ$, do not prevent the formation of unquenched orbital moments.  
This finding significantly increases the number of potential candidates for rare-earth-free hard-magnetic materials.

%The question remains on why the coercivity observed in 
%estimate anisotropy field based on calculated M_sat = sqrt2*2.2 - 2*2.2. dann vergleich mit Li3N
%For this reason we refrain from easy-plane - better easy axes in the plane

%%%%%%%%%%%%%%%%%%%%%%%%%%%%%%%%%%%%%%%%%%
\section{Materials and Methods}
Starting materials were lithium rod (Evochem, 99\,\%), iron powder (Alfa Aesar, 99.998\,\%), strontium nitride (Sr$_2$N) powder [prepared from strontium metal (Alfa Aesar, distilled dentritic pieces, 99.8\,\%) and nitrogen (Praxair, 99.999\,\%, additionally purified by molecular sieves)], and sodium azide (NaN$_3$) powder (Roth, 99\,\%) as a further nitrogen source. 
Tantalum ampules were produced on-site from pre-cleaned tantalum tube and tantalum sheet in an arc furnace located within a glove box. 

Laboratory powder X-ray diffraction data of finely ground (gray) powder samples were collected on a Huber G670 imaging plate Guinier camera using a curved germanium (111) monochromator and Cu-$K\alpha$1 radiation in the range $4^\circ \leq 2\theta \leq 100^\circ$ with an increment of $0.005^\circ$ at 293(1)\,K.
The powder samples were placed between Kapton foils to avoid degradation in air. Preliminary data processing was done using the WinXPow program package\,\cite{stoe}. 
Rietveld refinement of the structure of Li$_2$Sr[(Li$_{1-x}$Fe$_x$)N]$_2$ was performed with the software package Jana2006\,\cite{Petricek2014}. 
Small intervals of the diffraction pattern between $2\theta = 20^\circ{\text -}25^\circ$ corresponding to amorphous degradation products were excluded during the refinement. 
After background correction, profile and lattice parameters were refined by using Pseudo-Voigt profile functions and Berar-Baldinozzi's asymmetry model before refining atomic positions and isotropic thermal displacement factors. 

The crystal structure and the composition of the title compound Li$_2$Sr[(Li$_{1-x}$Fe$_x$)N]$_2$ was refined from single-crystal X-ray diffraction data which were collected at room temperature on a Rigaku AFC7 four circle diffractometer equipped with a Mercury-CCD detector (Mo-$K\alpha$ radiation, graphite monochromator). 
After data collection the structures were solved by direct methods, using SHELXS-97\,\cite{Sheldrick1997} and subsequently refined by using the full-matrix least-squares procedure with SHELXL-97\,\cite{Sheldrick1997b}.
Further details on the crystal structure investigations may be obtained from the Fachinformationszentrum Karlsruhe, 76344 Eggenstein-Leopoldshafen, Germany 
(Fax: +49-7247-808-666; e-mail: crysdata@fiz-karlsruhe.de), on quoting the depository number CSD-xxx, the names of the authors, and the journal citation.

The morphology of the Li$_2$Sr[(Li$_{1-x}$Fe$_x$)N]$_2$ sample and its metal composition were investigated using a scanning electron microscope Philips XL30 equipped with a Bruker Quantax EDX-System (Silicon drift detector, LaB$_6$ cathode). 
The EDX data were processed using the Esprit-Software.
The composition of the samples was further analyzed by inductively coupled plasma optical emission spectroscopy (ICP-OES) using a Varian Vista-MPX.
To this extent the samples were dissolved in dilute hydrochloric acid solution (4\,ml of 37\,\% hydrochloric acid added to 46\,ml deionized water).
Laue back reflection pattern were taken with a digital Dual FDI NTX camera manufactured by Photonic Science (tungsten anode, $U = 20$\,kV).
The magnetization was measured using a 7\,T Magnetic
Property Measurement System (MPMS3) manufactured by Quantum Design.

%Materials and Methods should be described with sufficient details to allow others to replicate and build on published results. Please note that publication of your manuscript implicates that you must make all materials, data, computer code, and protocols associated with the publication available to readers. Please disclose at the submission stage any restrictions on the availability of materials or information. New methods and protocols should be described in detail while well-established methods can be briefly described and appropriately cited.

%%%%%%%%%%%%%%%%%%%%%%%%%%%%%%%%%%%%%%%%%%
\vspace{6pt} 

%%%%%%%%%%%%%%%%%%%%%%%%%%%%%%%%%%%%%%%%%%
\acknowledgments{The authors thank Dr. U. Burkhardt, P. Scheppan, and S. H\"uckmann for experimental assistance. A. Mohs, A. Herrnberger and K. Wiedenmann are ackowlegded for technical support. This work was supported by the Deutsche Forschungsgemeinschaft (DFG, German Research Foundation) - Grant No. JE 748/1.}

%%%%%%%%%%%%%%%%%%%%%%%%%%%%%%%%%%%%%%%%%%
\authorcontributions{P.H. grew the single crystals and analyzed powder and single crystal X-Ray diffraction data that were collected by Y.P. T.B. performed the magnetization measurements and interpreted the data. M.F. collected and analyzed Laue-back-reflection data. A.J. wrote the paper with the help of all authors.}

%%%%%%%%%%%%%%%%%%%%%%%%%%%%%%%%%%%%%%%%%%
\conflictofinterests{The authors declare no conflict of interest.} 

\bibliographystyle{mdpi}
\bibliography{C:/Users/Administrator/Documents/presentations/paper/zitate}

%%%%%%%%%%%%%%%%%%%%%%%%%%%%%%%%%%%%%%%%%%
%% optional
\sampleavailability{Samples of the compounds are available from the authors.}

%%%%%%%%%%%%%%%%%%%%%%%%%%%%%%%%%%%%%%%%%%
\end{document}